# Novel Classification of Ischemic Heart Disease Using Artificial Neural Network

Giulia Silveri[1], Marco Merlo[2], Luca Restivo[2], Gianfranco Sinagra[2], Agostino Accardo[1]

[1]Department Engineering and Architecture, University of Trieste, Trieste, Italy
[2]Cardiovascular Department, University of Trieste, Trieste, Italy

**Abstract**

*Ischemic heart disease (IHD), particularly in its chronic stable form, is a subtle pathology due to its silent behavior before developing in unstable angina, myocardial infarction or sudden cardiac death. Machine learning techniques applied to parameters extracted form heart rate variability (HRV) signal seem to be a valuable support in the early diagnosis of some cardiac diseases. However, so far, IHD patients were identified using Artificial Neural Networks (ANNs) applied to a limited number of HRV parameters and only to very few subjects. In this study, we used several linear and non-linear HRV parameters applied to ANNs, in order to confirm these results on a large cohort of 965 sample of subjects and to identify which features could discriminate IHD patients with high accuracy. By using principal component analysis and stepwise regression, we reduced the original 17 parameters to five, used as inputs, for a series of ANNs. The highest accuracy of 82% was achieved using meanRR, LFn, SD1, gender and age parameters and two hidden neurons.*

## 1. Introduction

Ischemic heart disease (IHD) is a pathological condition characterized by an imbalance between myocardial oxygen supply and demand, mainly due to coronary artery atherosclerosis.

This disease lasts for a long time in a stable form but suddenly can become unstable due to plaque rupture or erosion causing unstable angina, myocardial infarction or sudden cardiac death. Consequently, an early detection of this condition could help to reduce the mortality rate associated with this disease [1].

IHD is diagnosed by clinical assesment with typical symptoms such as angina and laboratory test (elevation of Troponin I), ECG repolarization abnormalities (T-wave and ST), wall motion abnormalities (hypokinesia or akinesia) with echocardiography, and finally confirmed, invasively, by coronary angiography [2]. In the preliminary stage, the electrocardiogram (ECG) monitoring is commonly used for detecting IHD. This non-invasive technique allows the examination of the rate variability of heartbeats (HRV) assessing the functionality of the autonomic nervous system and the cardiovascular autonomic regulation [3]. Moreover, IHD subjects showed different values of HRV parameters from normal subjects [4-6] showing lower values of time [4, 5] and frequency domain [6] parameters as well as of non-linear parameters [5]. However, this method is sensitive to noise and till now it may not provide an accurate and clear categorization of normal and IHD subjects.

Recently, machine-learning methods are a powerful tool in HRV analysis to analyze data and particularly to identify some cardiovascular disease. However, even if some authors developed cardiovascular disease classification systems based on artificial neural network using few of parameters extracted from HRV [7-9] and some clinical features [10,11], most of them used arrhythmia public databases in which IHD was not considered separately [7-9]. In particular, Obbaya et al. [8] discriminated normal subjects from patients with low heart rate variability such as with congestive heart failure or with myocardial infarction disease taken from PhysioBank Interbeat Interval Database. Linear (in time and frequency domains) and non-linear HRV parameters were analyzed and considered separately, obtaining an average classification rate of about 95%.

In addition, Acharya et al [7], using the MIT-BIH arrhythmia database and three non-linear HRV parameters such as spectral entropy, Poincaré plot and largest Lyapunov exponent applied to an artificial neural network, obtained an accuracy level of 80-85% to classify ischemic/dilated cardiomyopathy. Finally, Dua et al. [9] extracted non-linear features from HRV and using the first six most significant principal components classified 10 IHD and 10 normal subjects with an accuracy of 89.5%.

In order to assess the latter result on a large sample of subjects and to identify which features, applied to artificial neural network (ANN) techniques, could discriminate IHD patients with high accuracy, in this study we examined the performance of several ANNs applied to linear and non-linear parameters extracted from HRV. Principal component analysis and stepwise regression were used to dimensionally reduce the number of features preserving the variance. The results of these classifiers were compared with those of an ANN that had as input all the HRV parameters. The influence of age and gender was considered adding these features as inputs of the ANNs.

## 2. Materials and Methods

In this research, we analyzed 965 subjects consecutively enrolled from December 2016 to October 2018 at the Cardiovascular Department, ASUGI, Trieste. Of these, 681 (316 males, aged 62±15 and 365 females, aged 64±15) were normal subjects and 284 (222 males, aged 71±10 and 62 females, aged 76±10) suffered from IHD. The latter presented the symptoms required by the guidelines for the diagnosis and management of chronic coronary syndromes [2]. The diagnostic of the disease was supported by coronary angiography. Normal subject did not present peripheral artery disease, thyroid disorders nor history of myocardial revascularization. The study was performed according to the Declaration of Helsinki and all patients gave written consent. All subjects completed a 24h Holter recording at a sampling frequency of 200 Hz, using a three channels tracking recorder (Sorin Group, Italy). The RR intervals were automatically extracted from records by using SyneScope analysis software. Data were analysed by using proprietary MATLAB® (MathWorks, USA) program that examined segments of 300s each [12]. For substituting the identified artefacts, a cubic spline interpolation based on the normal RR intervals was applied and the new time sequence was resampled at 2 Hz.

Several linear parameters were evaluated in the time domain: meanRR, SDNN, RMSSD, NN50 and pNN50 [14]. Frequency analysis was based on the power spectral density (PSD) [14] that was computed using the periodogram method with the Hamming window.

The low (LF, from 0.04 to 0.15Hz) and high frequency (HF, from 0.15 to 0.40Hz) bands together with their ratio (LF/HF) were calculated. The normalized powers in the same bands (LFn and HFn, respectively) was obtained dividing LF and HF power values by their sum. Finally, some non-linear parameters, like those extracted from the Poincaré plot (SD1, SD2 and SD1/SD2), Fractal Dimension (FD) and beta exponent were evaluated. They allow the quantification of short and long-term variability of RR sequence [15], the fractal-like behavior of RR as well as the self-affinity and system complexity. FD was evaluated directly from the RR time series by using Higuchi's algorithm [16] while beta exponent was calculated as the slope of the regression line in the relationship between the ln(PSD) and ln(Frequency) [17]. For each subject, linear and non-linear parameters were calculated on each segment and averaged along the 24h.

Artificial Neural Network (ANN) represents a powerful tool able to recognize patterns, manage data and learn [18]. In this paper, a multilayer feedforward neural network was developed using the generalized back-propagation algorithm (BPA) in which the learning process aims to reduce the overall system error to a minimum. Since both principal component analysis and stepwise fit regression method allow dimensionality reduction of features while preserving as much of the variance in the input characteristics as possible, we classified IHD and normal subjects, developing and comparing different ANNs using as input three different combinations of parameters. In the first situation (a) the first principal components explaining at least the 90% of system variability were considered, in the second case (b) the parameters extracted by using stepwise analysis presenting significant p-values (p<0.05) were used and finally, in the third situation, utilized for comparison, all the 15 linear and non-linear parameters were considered (c). Age and gender were added to these parameters and all of them were normalized and used as ANN inputs. Several networks with a number of hidden nodes varying between 2 and 6 and with two output nodes (one for each kind of subjects) were tested for each combination of inputs. The training and test sizes were respectively 75% and 25% of the total number of data. In order to assess the influence of data included in the two sets, 100 ANNs for each of the three situations and of the 5 different number of hidden neurons were examined, randomly extracting from original data those for training and test phases. The performance of each network classifier was measured through accuracy (ACC), sensitivity (SEN), specificity (SPE) and precision (PRE) parameters. Finally, the receiver operating characteristics curve (ROC) was used to depict trade-offs between hit rate (sensitivity) and false alarm rate (1.0-specificity) and the area under the curve (AUC) was evaluated. The distribution of the performance indexes was examined and the ANN presenting the highest accuracy in each of the three situations was finally selected.

## 3. Results

The number of ANNs inputs changed following the type of analysis carried out. At first, principal component analysis was carried out considering all the fifteen HRV parameters together with age and gender and the five most significant principal components, which account for 92% of the variance of the dataset, were selected as inputs for the first ANN (ANN1). The application of the stepwise method to the same set of parameters included in the final multilinear model only five features, nominally meanRR, LFn, SD1, gender and age, used as inputs for the second ANN (ANN2). Finally, all the seventeen parameters were considered as input for the third artificial neural network (ANN3). The ANNs showing the highest accuracy in the three situations were those in which 4, 2 and 3 hidden neurons, respectively, were used.

Table 1 presents the maxima and the mean (±1SD) values of the classification performance parameters for the three ANNs in the test phase. Among the three different neural network structures, ANN3 presented the lowest performance values while the other two classifiers had similar values with slightly higher values in ANN2 (ACC max=82%).

Table 1. Mean (±1SD) and maxima values of the classification performance indexes for the three ANNs in the test phase. SEN: sensitivity, SPE: specificity, PRE: precision, ACC: accuracy, AUC: area under the ROC curve.

|  |  | ANN1 | ANN2 | ANN3 |
|---|---|---|---|---|
| SEN (%) | Max | 52 | 58 | 49 |
|  | Mean±SD | 46.6±5.9 | 47.5±9.5 | 44.1±5.9 |
| SPE (%) | Max | 92 | 90 | 87 |
|  | Mean±SD | 46.0±3.1 | 86.3±3.6 | 84.4±3.4 |
| PRE (%) | Max | 74 | 69 | 61 |
|  | Mean±SD | 58.6±6.1 | 59.6±5.3 | 54.1±6.3 |
| ACC (%) | Max | 80 | 82 | 77 |
|  | Mean±SD | 74.4±2.5 | 74.7±2.6 | 72.5±2.5 |
| AUC (%) | Max | 83 | 85 | 80 |
|  | Mean±SD | 78.1±2.8 | 77.78±7.5 | 74.0±3.4 |

Figure 1 shows the distribution of the sensitivity, specificity, precision and accuracy obtained using 100 different combinations of data for training and test sets in the ANN1 and ANN2 with 4 and 2 hidden neurons, respectively. All distributions showed a somewhat symmetrical shape whose mean and maximum values are reported in Table 1.

Figure 2 shows the ROC curves obtained for the best ANNs in the three cases; the corresponding AUC values are reported in Table 1.

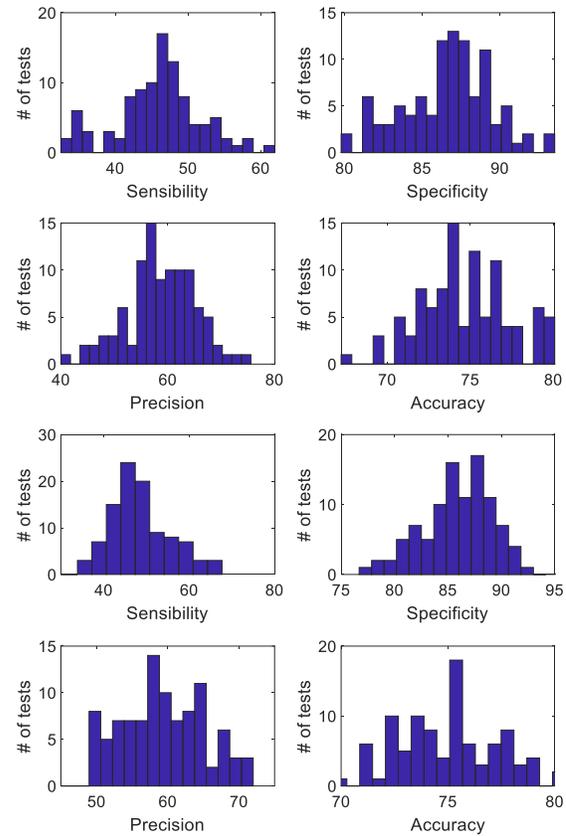

Figure 1. Distributions of the sensitivity, specificity, precision, accuracy obtained in 100 repetitions of the test set for ANN1 (top panels) and ANN2 (bottom panels).

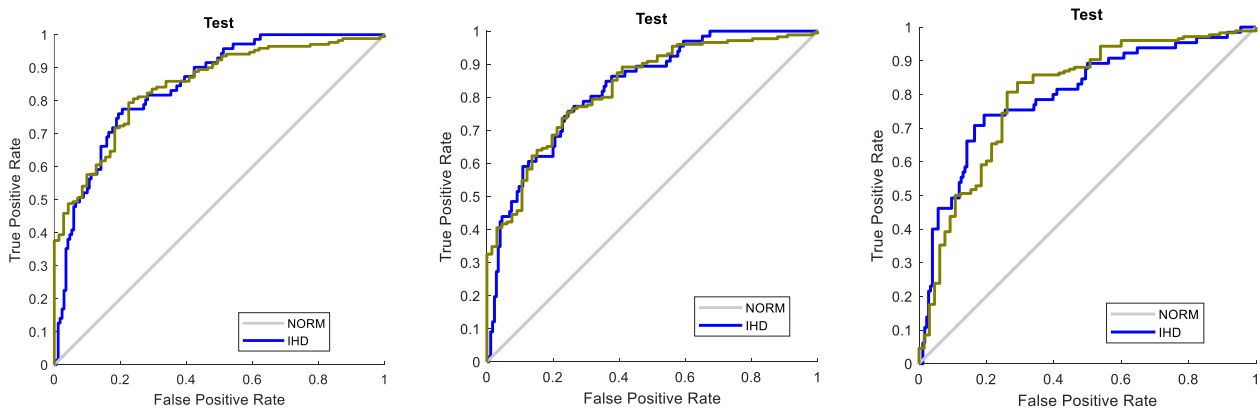

Figure 2. ROC curves in the test phase for the ANN1 (left), ANN2 (center) and ANN3 (right) producing the best accuracy.

## 4. Discussion

Cardiovascular diseases reduce the heart rate variability and, in particular, patients presenting ischemic heart disease showed significant lower linear and non-linear HRV parameters than normal subjects, due to an increase of sympathetic tone and/or a decreased of parasympathetic tone [4-6]. Recently, machine-learning techniques have become a useful research diagnostic tool for physician in the analysis of cardiovascular diseases [7-11], but until now, only Dua et al. [9] used these techniques to identify ischemic heart disease. They analyzed the HRV signals of 20 subjects applying principal component analysis to non-linear HRV parameters and obtaining an accuracy of 89.5%.

In this paper, we evaluated in a large cohort of subjects the performance of several ANNs, selecting the number of hidden neurons producing the greatest accuracy and the features obtained by using either principal component or stepwise regression analysis that allow the reduction of the inputs (from 17 to 5), decreasing the complexity of the model. Among the possible ANNs in which we considered the first five components obtained using PCA, the best network presented an accuracy of 80% and an AUC of 83% in the validation test. Although with a slightly lower performance, our network confirmed and extended the results of Dua et al. [9]. The difference between the accuracies could be due to the very small number of IHD and normal subjects considered in the previous work (only 20 subjects) in comparison with the large set (965 subjects) examined in this work and to the different ways of using data. Dua et al. [9] used repeated measurements on the same subjects while only independent measurements were considered in our work.

Higher accuracy (82%) and AUC (85%) values were reached by the ANN2 scheme with only five parameters as input. The network in which all the parameters were used as input (ANN3) showed the lowest performance. Low accuracy suggested that a pre-selection of features is needed to better discriminate IHD and normal subjects.

The proposed technique demonstrates the powerful capability of selected linear and nonlinear features and of ANN technique in differentiating, in a non-invasive way, normal from IHD patients. Our future work will examine other machine learning techniques such as decision tree or single vector machine applied to pre-selected parameters to evaluate their ability to discriminate normal and pathological subjects.

## Acknowledgments

Work partially supported by Master in Clinical Engineering, University of Trieste.

Address for correspondence:

Giulia Silveri
University of Trieste
Via A.Valerio 10, 34127, Trieste, Italy
giulia.silveri@phd.units.it